\magnification=1100

\hsize 17truecm
\vsize 23truecm

\font\twelvec=msbm10 at 10pt
\font\sevenc=msbm10 at 7pt
\font\fivec=msbm10 at 5pt

\newfam\co
\textfont\co=\twelvec
\scriptfont\co=\sevenc
\scriptscriptfont\co=\fivec

\def\det{\mathop{\rm det}\nolimits}
\def\exp{\mathop{\rm exp}\nolimits}

\def\id{\mathop{\rm Id}\nolimits}
\def\im{\mathop{\rm Im}\nolimits}

\def\ker{\mathop{\rm Ker}\nolimits}
\def\lim{\mathop{\rm lim}\nolimits}

\def\re{\mathop{\rm Re}\nolimits}
\def\spec{\mathop{\rm Spec}\nolimits}

\def\supp{\mathop{\rm supp}\nolimits}

\def\neigh{\mathop{\rm neigh}\nolimits}
\def\Op{\mathop{\rm Op}\nolimits}

\def\Sum{\displaystyle\sum}
\def\e{\mathop{\rm \varepsilon}\nolimits}

\baselineskip 15pt

\centerline{\bf HYPERBOLIC HAMILTONIAN FLOWS}

\centerline{\bf AND THE SEMI-CLASSICAL POINCAR\'E MAP}
\bigskip
\centerline{H.FADHLAOUI ${}^{2}$, H.LOUATI ${}^{1,2}$ {\it\&} M. ROULEUX ${}^{1}$}
\bigskip
\centerline {${}^{1}$ Universit\'e du Sud Toulon-Var, and Centre de Physique Th\'eorique}

\centerline {CPT, Case 907, 13288 Marseille Cedex 9, France}

\centerline {${}^{2}$ Universit\'e de Tunis El-Manar, D\'epartement de Math\'ematiques, 1091 Tunis, Tunisia}

\centerline{e-mail: rouleux@univ-tln.fr}
\bigskip
\noindent {\bf Abstract}: We consider semi-excited resonances created by a periodic orbit of hyperbolic type for Schr\"odinger type operators
with a small ``Planck constant''.
They are defined within an analytic framework based on the semi-classical quantization of Poincar\'e map in action-angle variables.
\medskip
\noindent {\bf 1. Introduction and main hypotheses}
\smallskip
Let $M$ be a Riemannian manifold. We wish to construct quasi-modes associated with a periodic orbit $\gamma_0$ for a $h$-PDO $H^w(x,hD_x;h)$
on $L^2(M)$ with Weyl symbol
$$H(x,\xi;h)=H_0(x,\xi)+hH_1(x,\xi)+h^2H_2(x,\xi)+\cdots\leqno(1.1)$$
This problem has a rich history, starting from the pionnering works of [Ba], [BaLaz] followed by [GuWe],
in case of a periodic orbit of elliptic type for the geodesic flow
on $M$. For more general Hamiltonians this was done by [Vo1] and [Ra].
The success of the method consisting in making the so-called ``Gaussian beams ansatz'', or the ``beam optics approximation'',
relies on the fact that quasi-modes are microlocalized on $\gamma_0\subset\{H_0=0\}$, 
or more precisely on the family $\gamma_E$ of elliptic
periodic orbits belonging to nearby energy surfaces $\{H_0=E\}$ (energy level reference is set to 0). 

In trying to extend this method to the case of a periodic
orbit of hyperbolic type, one is faced with the difficulty that a quasi-eigenfunction concentrated near such a $\gamma_0$ 
would have a complex eigenvalue.
On the other hand, because of local energy decay,
there is no invariant density on the stable or unstable manifolds $J^\pm$ associated
with the family $\gamma_E$. This prevents the existence of quasi-modes concentrated on $J^\pm$ as well. Nevertheless it is still possible to
derive semi-classical Gutzwiller-type trace formulas microlocally near $\gamma_0$
(i.e. by averaging over energy), whenever $\gamma_0$ is elliptic or hyperbolic, see [SjZw]. 

The problem of constructing quasi-modes in the hyperbolic case disappears in the framework of {\it resonances} instead, 
allowing for complex eigenvalues,
provided the Hamiltonian flow escapes to infinity for data outside the trapped set $K_E$ at energy $E$; periodic orbits $\gamma_E$
are actually trapped sets of simplest type. 

Resonances associated with hyperbolic orbits in the context of 
the wave equation outside some convex obstacles were considered in [Ik] and [G\'e], and later extended to the case of $h$-PDO 
in [Vo2], [G\'eSj], [Sj2] and [NoSjZw]. The latter paper deals not only with a periodic orbit, but also with
more general hyperbolic flows with compact $K_E$, provided it is topologically one dimensional. 

In this work, we are interested in the situation $K_0=\gamma_0$ already considered in [G\'eSj],
considering higher excited states, and allowing for a center manifold. Actually, our constructions rely
heavily on [Sj] and [NoSjZw].

Let us formulate now the problem more precisely. 
Let $M$ be a real analytic manifold, 
$H(x,hD_x;h)$ a $h$-PDO self-adjoint on $L^2(M)$ with real analytic coefficients, and Weyl symbol of the form (1.1). 
We think here especially of $M={\bf R}^n$, but resonances can be also defined on more general $M$, e.g. Riemannian
manifolds with negative curvature (see examples below).
The principal symbol $H_0$ is the corresponding classical Hamiltonian, and $H_1$ the sub-principal symbol. Assume 
$$H_0 \ \hbox{is of  principal type, i.e.} \ X_{H_0}(x,\xi)\neq0 \ \hbox{on} \ H_0(x,\xi)=0\leqno (1.2)$$ 
where $X_{H_0}$ denotes the Hamilton vector field. This means that $X_{H_0}$ has no fixed point on $\Sigma_0=\{H_0(x,\xi)=0\}$,
and hence on nearby energy surfaces $\Sigma_E=\{H_0(x,\xi)=E\}$.
Write $H=H_0$ for short, so long we are only interested on the classical Hamiltonian. Define the trapped set at energy $E$ as : 
$$K_E=\{\rho=(x,\xi)\in T^*M: H(\rho)=E, \Phi^t(\rho)=\exp tX_H(\rho) \ \hbox{is bounded for all real} \ t\}$$
and assume 
$$K_0=\gamma_0 \ \hbox{is a periodic orbit of period} \ T_0, \ \hbox{and} \ \gamma_0 \ \hbox{is of hyperbolic type}\leqno(1.3)$$ 
where hyperbolic means {\it partially hyperbolic}, i.e. {\it some} eigenvalue of Poincar\'e map is of modulus greater than 1.
Let us discuss this in more detail. 
Let $\Sigma=\Sigma_0$ be a Poincar\'e section (or {\it cross section}), i.e. a smooth hypersurface of $H^{-1}(0)$, transverse to $\gamma_0$,
$\{\rho_0\}=\Sigma_0\cap\gamma_0$, and  ${\cal P}_0:\Sigma_0\to\Sigma_0$ the  
Poincar\'e map, or {\it first return} map, with ${\cal P}_0(\rho_0)=\rho_0$.
Eigenvalues $\lambda_j$ of the real (linear) symplectic map $A=d{\cal P}_0(\rho_0)$ have been classified by Williamson; they
come in inverse and conjugate pairs, including possibly $\pm1$; the space ${\bf C}^{2(n-1)}$ has decomposition into symplectically
orthogonal spaces $E_1,E_{-1},E_{\lambda_j}\oplus E_{1/\lambda_j}$, with $E_{\pm1}$ symplectic, $E_{\lambda_j},E_{1/\lambda_j}$
isotropic for $\lambda_j\neq1,-1$ [here we do not count $\pm1,\lambda_j,1/\lambda_j$ with their multiplicity, but as distinct 
eigenvalues]. We know that $E_\pm$ and more generally,
$E_{\lambda}$ with real $\lambda$ are invariant by complex conjugation $\Gamma:(x,\xi)\mapsto\overline{(x,\xi)}$, and if $\lambda$ is not
real, either (1) $|\lambda|\neq1$ and $\overline\lambda,1/\lambda,1/{\overline\lambda}$ are also eigenvalues, and 
$E_\lambda\oplus E_{1/\lambda}\oplus E_{\overline\lambda}\oplus E_{1/{\overline\lambda}}$
is the complexification of a real symplectic space; or (2) $|\lambda|=1$ and $E_\lambda\oplus E_{1/\lambda}$ is the complexification
of a real symplectic space. Since we need to define the logarithm of $A$, we shall assume
$$E_{\pm1}=\{0\}, \ \hbox{and} \ E_\lambda=\{0\}, \ \forall \lambda\leq0\leqno(1.4)$$
though negative eigenvalues $\lambda\leq-1$ might occur if the center manifold is not orientable
(see [Sj2]). These eigenvalues are called {\it Floquet multipliers};
these with $|\lambda|=1$ are called {\it elliptic},
those with $|\lambda|\neq1$ {\it hyperbolic}. 

Provided (1.4) holds, it can be shown [IaSj] that we can make sense of $B=\log A$
as a real matrix, anti-symmetric with respect to the symplectic 2-form $\sigma$, i.e. ${}^\sigma B+B=0$. Complex eigenvalues $\mu=\mu(\lambda)$ of $B$
can be chosen in such a way that $\mu(\overline\lambda)=\overline{\mu(\lambda)}$. 
They are called {\it Floquet exponents}.
Correspondingly, 
these Floquet exponents with $\re\mu(\lambda)=0$ 
($|\lambda|=1$) are called {\it elliptic}, and those with $\re\mu(\lambda)\neq0$ ($|\lambda|\neq1$) {\it hyperbolic}. 
Under (1.4), eigenvalues of $B$
are of the form $\mu_j,-\mu_j\neq0$, with equal multiplicity $>0$, and so we can arrange so that all $\mu_j$ are distinct 
either with $\re \mu_j>0$ or $\re \mu_j=0$ and $\im \mu_j>0$. Let $r\leq n-1$ is the number of distinct $\mu_j$. 
We denote by $b$ the quadratic form $b(\rho)={1\over2}\sigma(\rho,B\rho)$. 

For simplicity, we also assume that $b$ is diagonalizable, 
which is the case when $r=n-1$. Denote by $F_{\mu_j}$ the eigenspace corresponding to $\mu_j$. 
It is possible to recombine the decomposition of ${\bf C}^{2(n-1)}$ as above into unstable $F^+$ or stable spaces $F^-$; namely let
$F^+=\oplus_{j=1}^rF_{\mu_j}$, $F^-=\oplus_{j=1}^rF_{-\mu_j}$. Then $F^\pm\approx{\bf C}^{(n-1}$ are (complex)
Lagrangian subspaces of ${\bf C}^{2(n-1)}$
invariant under the Hamiltonian flow $X_b$, and 
if $\theta>0$, $e^{-i\theta}X_b$ is ``expansive'' on $F^+$, ``contractive'' on $F^-$. We know also that $F^\pm$ are
{\it dissipative} in the sense ${1\over2i}\sigma(u,\overline u)>0$ for all $u\in F^\pm$, $u\neq0$. 
They are known as the linear (reduced) {\it Maslov germ}. This decomposition of ${\bf C}^{2(n-1)}$ into invariant subspaces holds for $A=e^B$ as well.
In a diagonal, real basis, $b(\rho)$ assumes the form of a linear combination of 
elementary quadratic polynomials. When all $\mu_j$ belong to the imaginary axis (``completely elliptic'' case)
the corresponding symplectic coordinates are known as {\it Poincar\'e coordinates}, or {\it harmonic oscillator coordinates}.
The {\it partial hyperbolicity} assumption reads
$$\exists j\in\{1,\cdots,r\}, \ \re\mu_j>0\leqno(1.5)$$
The {\it hyperbolic dimension} is defined as the number of $j$'s such that (1.5) holds.
We make also the non-resonance assumption:
$$\forall k_1,\cdots,k_{r}\in{\bf Z}: \ \Sum_{j=1}^r k_j\mu_j\in2i\pi{\bf Z}\Longrightarrow\Sum_{j=1}^r k_j\mu_j=0\leqno(1.6)$$
Note that when $n=2$, and $\mu_1$ elliptic, the non-resonance assumption just states that $i\mu_1$ (the rotation number on Poincar\'e
section) is irrational. 
For later purposes, we introduce the additional {\it strong non-resonance assumption} $r=n-1$ and:
$$\forall k_1,\cdots,k_{n-1}\in{\bf Z}: \ \Sum_{j=1}^{n-1} k_j\mu_j\in2i\pi{\bf Z}\Longrightarrow k_j=0, \ j=1,\cdots,n-1\leqno(1.7)$$
Poincar\'e map is actually related to the variational problem as follows: 
let $H''(\rho)$ be the Hessian of $H$ and ${\cal J}$ the symplectic matrix. Then 
for all $\rho_0\in\gamma_0$, $\spec({\cal J}H'')(\rho)=\{0\}\cup\{\pm\mu_j, 1\leq j\leq r\}$.  
Hypothesis $E_1=\{0\}$ (see (1.4)) means that 0
has multiplicity 2 in $\spec({\cal J}H'')(\rho)$, or equivalently the kernel of the differential of the symplectic map 
${\cal P}_0=\exp T_0X_H$ at $\rho_0$
consists only in the Hamiltonian vector field $X_H(\rho_0)$. This makes together with (1.5) the {\it non degeneracy condition}, which
guarantees that for $|E|\leq\e _0$, there is a one-parameter family of periodic orbits $K_E=\gamma_E\subset H^{-1}(E)$,
all of hyperbolic type, all transverse to $\Sigma$, their period $T(E)$ depending analytically on $E$. 
The corresponding Poincar\'e maps ${\cal P}_E:\Sigma_E\to\Sigma_E$ enjoy the same properties
as ${\cal P}_0$, with same hyperbolic dimension.

One could try to take right away
$H$ to its ``Birkhoff normal form'' microlocally near $\gamma$; this is a strategy followed in [Sj2] when $n=2$. For higher dimensions,
we might use the reduction of a system with periodic coefficients to a normal form, see [ArKoNe,Sect.8.4.2] and references therein, and 
then quantize it;  
instead we construct the monodromy operator ${\cal M}(z)$ associated with $H(x,hD_x;h)$ as in [NoSjZw], also for complex energies $z$
with $|\im z|={\cal O}(h^\delta)$, $0<\delta\leq1$. This reflects the physical fact that Poincar\'e map actually carries most of relevant information
in such a dynamical system. 
\medskip
\noindent {\bf 2. Microlocal reduction of $H$ near $\gamma$, quantum monodromy and the Grushin operator}
\smallskip
Under Hypothesis (1.2)-(1.3) on $H$ and (1.4)-(1.5) on the linear flow (i.e. the flow of $X_b$) we shall define
${\cal M}(E)$ as a $h$-FIO which quantizes Poincar\'e map.

To start with, we proceed somewhat formally, ignoring the analytic dilations and Lagrangian deformations
required to define resonances in a proper way.
As $H(x,hD_x;h)$ is of real principal type, for all $\rho\in\gamma_0$, there is a local canonical transformation $\kappa_U:T^*{\bf R}^n\to T^*M$
defined near $((0,0),\rho)$ and a (microlocally)
unitary FIO $U:L^2({\bf R}^n)\to L^2(M)$ with canonical relation $C'_U=\{(x,\xi;y,-\eta): (y,\eta)=\kappa_U(x,\xi)\}$ such that $U^{-1}HU=hD_{x_n}$ 
(all equalities between operators are understood in the microlocal sense, and modulo
smoothing operators, the regularity being clear from the context; we refer to [Iv] for the general calculus of $h$-PDO's and $h$-FIO's.~) 
Thus, Poincar\'e section near $\rho$ can be defined by
$\Sigma=\{(y,\eta)=\kappa_U(x',0,\xi',0), (x',\xi')\in\neigh(0,0)\}$. We set $\widetilde\Sigma=(\kappa_U)^{-1}(\Sigma)$ that 
we identify locally with ${\bf R}^{2(n-1)}$. 
Let $\ker _{\rho}(H-z)=U(\ker (hD_{x_n}-z))$, and identify $\ker _{\rho}(H-z)$ with $L^2({\bf R}^{n-1})$ through Poisson
operator ${\cal K}(z)v(x')=e^{izx_n/h}v(x')$, $(x',x_n)\in\widetilde\Omega$. For the corresponding Poisson operator on $M$,
microlocalized on a neighborhood $\Omega$ of $\rho$ (such that $|x_n|\leq2\e $ in $\supp(\kappa_U)^{-1}(\Omega)\subset\widetilde\Omega$) we take 
$K(z)=\chi^w_\Omega U{\cal K}(z)$, where $\chi_\Omega$ is a smooth cut-off equal to 1 near $\Omega$. It is readily seen to be a $h$-FIO
such that $(H-z)K(z)=0$ in $\Omega'\subset\subset\Omega$, and propagating in the ``backward'' direction.

Let also $\widetilde\chi\in C^\infty({\bf R})$ be a smooth step function, $\widetilde\chi(x_n)=0$
for $x_n\leq-\e /2$, $\widetilde\chi(x_n)=1$ for $x_n\geq\e /2$. Since the commutator ${i\over h}[hD_{x_n},\widetilde\chi]=
\widetilde\chi'(x_n)$ is localized in the region
of the step we have the normalization property 
$$({i\over h}[hD_{x_n}, \widetilde\chi^w]{\cal K}(z)v_1|{\cal K}(\overline z)v_2)_{L^2({\bf R}^n)}=(v_1|v_2)_{L^2({\bf R}^{n-1})}$$
which is independent of the precise choice of $\widetilde\chi$ as above. 
We shall denote by ${\cal H}(\widetilde\Sigma)$ the space of $L^2$ fonctions microlocalized on 
$\widetilde\Sigma$, and similarly with $\Sigma, \Omega$, etc\dots 
This normalisation carries to ${\cal H}(\Sigma)$
(the set of transversal data) through
$$({i\over h}[H, \chi^w]K(z)v_1|K(\overline z)v_2)_{L^2(M)}=(v_1|v_2)_{L^2}$$
where the symbol $\chi$ is such that $\chi^w=U\widetilde\chi^wU^*$. Thus the LHS defines (microlocally) 
a norm on ${\cal H}(\Sigma)$, called the {\it quantum flux norm}, and we have
$$K(\overline z)^*{i\over h}[H,\chi^w]K(z)=\id \ \hbox{on} \ {\cal H}({\Sigma})$$
defining the microlocal inverse $R_+(z)={\cal K}(\overline z)^*U^*\chi^w_\Omega{i\over h}[H,\chi_f^w]$ of $K(z)$,
where the subscript $f$ has been added to underline the fact that $R_+(z)$ propagates singularities in the ``forward'' direction.
Similarly, we introduce $R_-(z)={i\over h}[H,\chi_b^w]K(z)$ acting on transversal data, and $\chi_b$ is defined 
as $\chi_f$, with $\chi_b=1$ on a neighbhd of $\supp\chi_f$. This is again a $h$-FIO propagating in the ``backward'' direction.

We next define {\it monodromy operators} acting on transversal data. 
Following [NoSjZw] it may be convenient to introduce another Poincar\'e section, so 
we take $\rho_1=\exp(T_0/2)X_{H}(\rho_0)$ $\in\gamma_0$
and $\Sigma_1$ the Poincar\'e section passing through $\rho_1$; we label henceforth by subscript
$j=0,1$ all the local objects defined before belonging to $\rho_j$, adding $E$ as an argument if we want to stress also they belong
to energy $E$. 

Starting with $v_0\in {\cal H}(\widetilde\Sigma_0)$, we apply Poisson operator, which we continue as a $h$-FIO all along $\gamma_0$, 
in a neighborhood $\Omega^+_0$, $\Omega_0\subset\Omega^+_0$, of a segment $\gamma_0$ forward of $\rho_0$, 
namely
$K^+_0(z)=\chi^w_{\Omega^+_0} U_0{\cal K}(z)$, where $\chi_{\Omega^+_0}$ is a cutoff extending $\chi_{\Omega_0}$ accordingly. 

We extend also $\chi_b$ from $\rho_0$ along $\gamma_0$
to a cutoff function $\chi_{b,0}$, with $\chi_{b,0}(\rho_1)=0$, and choose $\chi_{b,1}\in C_0^\infty(M)$ such that $\chi_{b,0}+\chi_{b,1}=1$
in a neighborhood of the segment $[\rho_0,\rho_1]\subset\gamma_0$. We do the same with $\chi_f$.
Near $\Sigma_1$, we truncate $K^+_0(z)v_0$ by $\chi_{b,0}^w$.
Thus $u_0=\chi_{b,0}^wK^+_0(z)v$ satisfies ${i\over h}(H-z)u={i\over h}[H,\chi_{b}^w]K^+(z)v_0$, which we decompose 
as the sum of $R^0_-(z)v_0$ near $\rho_0$ and ${i\over h}[H,\chi_{b,1}^w]K^+_0(z)v_0$. 
This defines a (partial) monodromy operator
${\cal M}_{10}(z)$ as $K^+_0(z)v_0=K_1(z){\cal M}_{10}(z)v_0$, microlocally near $(0,0)$. Explicitely 
$${\cal M}_{10}(z)=K_1(\overline z)^*{i\over h}[H,\chi_{f,1}^w]K^+_0(z)=R^1_+(z)K_0^+(z)$$
Moving further towards $\Sigma_0$ we define similarly the (partial) monodromy operator
${\cal M}_{01}(z)$. Both ${\cal M}_{10}(z)$ and ${\cal M}_{01}(z)$ are
$h$-FIO defined on $\widetilde\Sigma_0$ and $\widetilde\Sigma_1$ respectively, microlocally unitary for real $z$. The 
monodromy operator (or semi-classical Poincar\'e map)
${\cal M}_0(z)={\cal M}_{01}(z){\cal M}_{10}(z)$ is again a $h$-FIO, whose canonical relation is given by Poincar\'e map ${\cal P}_0$.

This operator contains most of information relative to the Grushin problem
near $\gamma_0$, that we formulate as follows [NoSjZw]. Consider (formally)
${\cal G}(z)=\pmatrix{{i\over h}(H-z)&R_-(z)\cr R_+(z)&0\cr}$, $R_-(z)(u_-^0,u_-^1)(x)=R_-^0(z)u_-^0(x)+R_-^1(z)u_-^1(x)$,
where $u_-^j(x')\in {\cal H}(\widetilde\Sigma_j)$, and $R_+(z)=\bigl(R_+^0(z),R_+^1(z)\bigr)$. This is a Fredholm operator of index 0, and its 
inverse is given by ${\cal E}(z)=\pmatrix{E(z)&E_+(z)\cr E_-(z)&E_{-+}(z)\cr}$, where $E_+(z)=\Sum_j\chi^j_bK^+_j(z)$, $E_-(z)=R_+(z)\widehat E(z)$,
$E_{-+}(z)=\pmatrix{\id _1&{\cal M}_{10}(z)\cr{\cal M}_{01}(z)&\id _0\cr}$, $\id _j$ is the identity of ${\cal H}(\widetilde\Sigma_j)$, and 
$\widehat E(z)=\int_0^\tau e^{-it(H-z)/h}\, dt$ is a forward parametrix, where $\tau>0$ is chosen small enough, say $\tau=4\e $. 
Thus, $\ker (H-z)\neq0$ iff $\ker(\id-{\cal M}_0(z))\neq0$. We don't specify the domain of these operators, since they will have to be
modified to take analytic distorsions and Lagrangian deformations into account.
\medskip
\noindent {\bf 3. Logarithm of ${\cal M}_0(z)$ and semi-classical LSNF of Poincar\'e map}
\smallskip
Here we construct the
``logarithm'' $P$ of ${\cal M}_0(z)$ as a $h$-PDO. Thanks to (1.7)
we shall then be able to take $P$ to its Birkhoff normal form (BNF) and ${\cal M}(E)$ to its Lewis-Sternberg normal form (LSNF).

Again we begin to ignore Lagrangian deformations (due to the fact that we deal with complex energies)
and proceed in a formal way. We work locally on $\widetilde\Sigma_0$.
Let ${\cal P}_0$ Poincar\'e map satisfy (1.3)-(1.6), ${\cal P}_0(\rho_0)=\rho_0$, 
$\mu_j$ its Floquet exponents, and $p_0(\rho)=b(\rho)$. As before, consider a local chart in $M$ where the cross section is $\Sigma$
and $\rho_0=0$. We know [IaSj,Thm 1.3] that there exists (a unique) $p\in C^\infty(\widetilde\Sigma;{\bf R})$ 
such that $p(\rho)=p_0(\rho)+{\cal O}(\rho^\infty)$ and ${\cal P}_0(0)=\exp X_p(\rho)+{\cal O}(\rho^\infty)$. Actually this holds for the family
${\cal P}_E$, $|E|$ small enough, $p$ depending smoothly on $E$, and such that $\rho\mapsto p(\rho,E)$ has a non critical degenerate point
at $\rho_E$.
Assume as before (for simplicity) the $\mu_j$ are semi-simple, 
i.e. $B$ diagonalizable.
The stable/unstable manifold theorem as is formulated in [Sj1] (see also [Ro2]) allows to construct the reduced complex Maslov germ from
its linear part. Namely, for $\theta>0$ is small enough, there are $X_p$-invariant complex Lagrangian
manifolds $\Lambda_\pm$ passing through $\rho_0$, such that
$T_{\rho_0}(\Lambda_\pm) = F_\pm$. Within $\Lambda_+$ (resp. $\Lambda_-$), $\rho_0$ is repulsive (resp. attractive) for 
$e^{i\theta}X_H$, and $p|_{\Lambda_\pm}= 0$. We can also find complex symplectic coordinates $(\zeta,\zeta^*)$,
$\Lambda_+= \{ \zeta^*=0 \}, \ \Lambda_-=\{ \zeta=0\}$. 
In these coordinates $p(x,\xi) = \langle \widetilde B(\zeta,\zeta^*)\zeta,\zeta^*\rangle$
where $\widetilde B(\zeta,\zeta^*)$ is a $(n-1)\times(n-1)$ matrix with smooth coefficients
such that $d\widetilde B(\rho_0)=B$. For corresponding real symplectic coordinates, 
since $p(\rho)=b(\rho)+{\cal O}(|\zeta,\zeta^*|^3)$, we can split $p(x,\xi)$ as an ``elliptic'' 
(in the ``harmonic oscillator'' coordinates) plus an ``hyperbolic'' term. 

Having brought the canonical transform of ${\cal M}(z)$ to the form $\exp X_p(\rho)$, following [IaSj,Thm 3.2] we proceed to take ${\cal M}(z)$
itself to the form $e^{iP^w(x,HD_x;z,h)/h}$ (modulo smoothing operators), where the symbol $P$ of $P^w(x,hD_x;z,h)$ has principal part $p$. 
This is done by a deformation argument, starting again from the metaplectic operator associated with the linearized Poincar\'e map. 

At last we make use of the non-resonance condition (1.7) on Floquet exponents to bring 
$P^w(x,$ $hD_x; z,h)$ in its Birkhoff normal form, using $\zeta,\zeta^*$ coordinates. This is done as in [IaSj,Sect.4], the action variables
$\iota_j$, $j=1,\cdots,n-1$ being the elementary quadratic polynomials that build up $b(\rho)$. Thus there is a $h$-FIO $V(z)$ microlocally unitary
on ${\cal H}(\widetilde\Sigma_0)$ (for real $z$), and a symbol $F(\iota;z,h)\sim F_0(\iota;z)+hF_1(\iota;z)+\cdots$, such that 
$V(z)^{-1}{\cal M}(z)V(z)=\exp[-iF(\iota;z,h)/h]$. 

Since we will allow $z$ to take complex values, and all operators we have constructed so far depend on $z$, with 
$|\im z|={\cal O}(h^\delta)$ (which is the expected width of 
resonances), we need to consider their almost analytic extensions in the complex domain, with suitable growth. In particular, 
all cut-off functions above will have to be chosen in suitable Gevrey classes $G^s({\bf R}^n)$, with $1-{1\over s}<\delta$.
We refer to [Ro1] for the overall strategy, but here we need to extend deformations of $h$-PDO's to $h$-FIO's; see also [SjZw] and references therein
for the $C^\infty$ case.
\medskip
\noindent {\bf 4. Lagrangian deformations and resonances}
\smallskip
Remembering now that we actually deal with resonances, first we carry on $H^w(x,hD_x;h)$ an ``analytic dilation'' outside of the
trapped set $\gamma_0$ and apply the considerations above (without any change) to the ``dilated'' (non self-adjoint) operator $H^w_\theta(x,hD_x;h)$ 
for small $\theta$. We give below the strategy of the proof.

Resonances are actually the (complex) eigenvalues of $H^w_\theta(x,hD_x;h)$, which we need to make (microlocally) elliptic 
outside of the trapped set $K_E=\gamma_E$ by introducing
some local $L^2$ weighted spaces. 
Taking (1.3) into account implies that outside any neighbhd of $\gamma_E$, 
there exists an {\it escape function}, i.e. a smooth function
that grows along the Hamiltonian flow $\Phi^t(\rho)=\exp tX_H(\rho)$ (see[G\'eSj]). 

Because of the properties of the Grushin operator considered in Sect.2, the main task will be to extend ${\cal M}_0(z)$ to some local 
$L^2$ weighted spaces, and construct first
an escape function on the Poincar\'e section near $\rho_E$ for the ``logarithm''$P^w_\theta(x,hD_x;z,h)$ of ${\cal M}(z)$.
In other words, we put a weight on ${\cal H}(\widetilde\Sigma_0)\times {\cal H}(\widetilde\Sigma_0)$, 
so that $E_{-+}(z)=-\id _0+{\cal M}_0(z)$, 
acting on the corresponding space, will have the right behavior. 

As in [Sj1], [KaKe] we can construct $g(\rho)$, $\rho\in\Sigma_0$, a ``conjugate'' quadratic function to 
the principal symbol $p=p_0$ of $P(x,hD_x;z,h)$,  
that depends only on the action variables $\iota''$ corresponding to eigenvalues $\mu_j$
with $\re \mu_j>0$, namely $X_pg(\rho)\sim|\rho|^2$ on $p(\rho,E)=0$.
For $s>1$, define the ``conjugate function'' $g_s(\iota'';h)=h^{1-1/s}g(\iota'')$, and the corresponding ``conjugate operator'' 
defined (formally) as $W=\Op  e^{g_s(\iota'')/h}$. We denote by ${\cal H}_g(\widetilde\Sigma_0)\times {\cal H}_g(\widetilde\Sigma_0)$ the weighted space.
Baker-Campbell-Hausdorff formula then shows that $W^{-1}e^{iP/h}W$ is of the form
(modulo smoothing operators) $e^{iP_g/h}$ where $P_g$ has ``leading symbol'' $p_g=p+iX_pg_s$. This is a $h$-FIO with complex phase,
acting naturally on polynomials in the action variables $\iota$ , and whose eigenfunctions are (formally) of the form $\iota^\alpha$,
$\alpha\in{\bf N}^{n-1}$. 

Similarly, we analyse the action of $E_+(z;h)$ on ${\cal H}_g(\widetilde\Sigma_0)\times{\cal H}_g(\widetilde\Sigma_0)$. We consider also 
${\cal H}({\cal T})$, where ${\cal T}$ is a thin ``tube'' around $\gamma_0$, and a corresponding weighted space ${\cal H}_G({\cal T})$, where $G$
is the flow out of $g$ through $X_H$. Operators $E(z;h)$ and $E_-(z;h)$ are again well-behaved on this space, and it follows easily that 
Grushin problem is well-posed near $\gamma_0$. Non trapping condition (1.3) also shows that this holds globally.  We can so far summarize our 
constructions, generalizing a result of [G\'eSj] for a smaller window
of the form $[-\e _0,\e _0]-i]0,Ch\log{1\over h}]$, namely:
\medskip
\noindent {\bf Theorem}: {\it Assume Hamiltonian (1.1) satisfies (1.2) to (1.7), and $0<\delta\leq1$. For $\e _0,C>0$ 
let ${\cal W}_h=[-\e _0,\e _0]-i]0,Ch^\delta]$. Then if $\e _0,C>0$ are small enough, there is $h_0>0$ small enough
and a family of matrices $N(z,h)=\Pi_h{\cal M}(z)\Pi_h+{\cal O}(h^N)$ of rank $\sim h^{-n(1-\delta)}$, such that  
the zeroes (with correct multiplicities)
of $\zeta(z,h)=\det(\id-N(z,h))$ give all resonances
of $H^w(x,hD_x;h)$ in
${\cal W}_h$ with correct multiplicities.}
\smallskip
These resonances, lying on ``strings'' in the lower-half complex plane, are labelled by a ``longitudinal quantum number'' 
(given by Bohr-Sommerfeld quantization condition along the periodic orbit) and 
``transversal quantum numbers'' $\alpha\in{\bf N}^{n-1}$ corresponding to the excitation modes on a cross-section.
The precise quantization rule, including the determination of Gelfand-Lidskii, or Conley-Zehnder indices, will be given elsewhere. 
\medskip
\noindent {\bf 5. Applications}
\smallskip
Hyperbolic periodic orbits occur in many physical examples:
(1) $H=-h^2\Delta+|r|^{-1}+ax_1$ on ${\bf R}^n$ (repulsive Coulomb potential perturbed by Stark effect) near an energy level
$E>2/\sqrt a$. (2) When $M$ is a 2-D Riemannian manifold there is a closed geodesic that minimizes the energy integral,
and a theorem of Poincar\'e (see [MoZh,p.169] says that its Floquet multipliers $\lambda$, $\lambda^{-1}$ are real 
(in higher dimensions, hyperbolic dimension $\leq1$ is the rule [A]). Our theory applies e.g. if $M$ is the one-sheeted hyperboloid 
in ${\bf R}^3$, and $\gamma_0$ its ``equatorial'' circle. 
(3) The ``hip-hop'' orbit [LeOffBuKo] in the $2N$ body problem with Newton potential is a hyperbolic periodic orbit. 
It would be suitable to extend this example to Coulomb-like potentials. 
\medskip
\noindent {\bf References}
\smallskip
\noindent [A] M.-C.Arnaud. On the type of certain periodic orbits minimizing the Lagrangian action. Nonlinearity 11, p.143-150, 1998.

\noindent [ArKoNe]  V.Arnold, V.Kozlov, A.Neishtadt. Mathematical aspects of classical and celestial mechanics. Encyclopaedia of Math. Sci.,
Dynamical Systems III, Springer, 2006.

\noindent [Ba] V.Babich Eigenfunctions concentrated near a closed geodesic [in Russian], Vol.9, Zapiski Nauchnykh Seminarov 
LOMI, Leningrad, 1968. 

\noindent [BLaz] V.M.Babich, V.Lazutkin.  Eigenfunctions concentrated near a closed geodesic. Topics in Math. Phys., Vol.2, M.Birman, ed. 
Consultants' Bureau, New York, 1968, p.9-18

\noindent [Ge] C.G\'erard. Asymptotique des p\^oles de la matrice de scattering pour 2 obstacles strictement convexes. M\'emoire Soc.
Math. France, S\'er.2 (31), p.1-146, 1988.

\noindent [GeSj] C.G\'erard, J.Sj\"ostrand. Semiclassical resonances generated by a closed trajectory of hyperbolic type.
Comm. Math. Phys. 108, p.391-421, 1987.

\noindent [GuWe] V.Guillemin, A.Weinstein. Eigenvalues associated with a closed geodesic. Bull. AMS 82, p.92-94, 1976.

\noindent [IaSj] A.Iantchenko, J.Sj\"ostrand. Birkhoff normal forms for Fourier integral operators II. American
J. of Math., 124(4), p.817-850, 2002.

\noindent [I] M.Ikawa. On the existence of poles of the scattering matrix for several convex bodies. Proc. Japan Acad. Ser.A Math. Sc., 64,
p.91-93, 1988.

\noindent [Iv] V.Ivrii. Microlocal Analysis and Precise Spectral Asymptotics. Springer-Verlag, Berlin, 1998.

\noindent [KaKe] N.Kaidi, Ph.Kerdelhue. Forme normale de Birkhoff et r\'esonances. Asympt. Analysis 23, p.1-21, 2000.

\noindent [LeOffBuKo] M.Lewis, D.Offin, P.-L.Buono, M.Kovacic. Instability of the periodic hip-hop orbit in the $2N$-body problem with equal masses.
Disc. Cont. Dyn. Syst. Vol.33(3), p. 1137-1155, 2013.

\noindent [MoZh] J.Moser, E.Zehnder. Notes on Dynamical systems. American Math. Soc., Courant Inst. Math. Sci. Vol.12, 2005.

\noindent [NoSjZw] S.Nonnenmacher, J.Sj\"ostrand, M.Zworski. From Open Quantum Systems to Open Quantum maps.
Comm. Math. Phys. 304, p.1-48, 2011

\noindent [Ra] J.V.Ralston. On the construction of quasi-modes associated with periodic orbits. Comm. Math. Phys. 51(3) p.219-242, 1976.

\noindent [Ro] M.Rouleux. {\bf 1}. Resonances for semi-classical Schr\"odinger operators of Gevrey type. 
Hokkaido Math. J., Vol.30 p.475-517, 2001. {\bf 2}. Semiclassical integrability, hyperbolic flows, and the Birkhoff
normal form. Canadian J. of Math. Vol.56 (5), p.1034-1067, 2004. 

\noindent [Sj] J.Sj\"ostrand. {\bf 1}. Semi-classical resonances generated by a non-degenerate critical point, {\it in} Lect.Notes in Math. Vol.1256,
Springer, p.402-429. {\bf 2}. Resonances associated to a closed hyperbolic trajectory in dimension 2. Asympt. Analysis 36, p.93-113, 2003.

\noindent [SjZw] J. Sj\"ostrand and M. Zworski. Quantum monodromy and semi-classical trace formulae, J. Math. Pure Appl.
81(2002), 1-33. Erratum: http://math.berkeley.edu/~zworski/qmr.pdf

\noindent [Vo] A.Voros. {\bf 1}. Semi-classical approximations. Ann. Inst. H.Poincar\'e, 24, p.31-90, 1976.
{\bf 2}. Unstable periodic orbits and semiclassical quantization. J.Phys. A(21), p.685-692, 1988. 
\bye